\documentclass{jnmauth}
\include{times}

\begin{document}
\JNM{43}{59}{17}{28}{04}

\runningheads{A.\ Aste et al.}
{Full Hydrodynamic Model}

\title{Full Hydrodynamic Simulation of GaAs MESFET's}

\author{Andreas Aste\affil{1}, R\"udiger Vahldieck\affil{2},
Marcel Rohner\affil{3}}

\address{\affilnum{1}\ Institute for Theoretical Physics,
Klingelbergstrasse 82, 4054 Basel, Switzerland\\
\affilnum{2}\
Laboratory for Electromagnetic Fields and Microwave Electronics,\\
Gloriastrasse 35, 8092 Z\"urich, Switzerland\\
\affilnum{3}\
Electronics Laboratory, Swiss Federal Institute of
Technology, 8092 Z\"urich, Switzerland}

\corraddr{Andreas Aste, Institute for Theoretical Physics,
Klingelbergstrasse 82, 4054 Basel, Switzerland}

\footnotetext{Work supported by the Swiss National
Science Foundation, Project no. 2100-57176.99}

\received{1 January 2003}
\revised{25 September 2003}
\accepted{1 November 2003}

\begin{abstract}
A finite difference upwind
discretization scheme in two dimensions
is presented in detail for the transient simulation
of the highly coupled non-linear partial differential equations of
the full hydrodynamic model, providing thereby a practical engineering
tool for improved charge carrier transport simulations at high electric
fields and frequencies.
The discretization scheme preserves the
conservation and transportive properties of the equations. 
The hydrodynamic model is able to describe inertia effects which
play an increasing role in different fields of micro- and optoelectronics,
where simplified charge transport models like the
drift-diffusion model and the energy balance model are no longer applicable.
Results of extensive numerical simulations are shown for a two-dimensional
MESFET device. A comparison of the hydrodynamic model to the commonly
used energy balance model is given and the accuracy of the results
is discussed.
\end{abstract}

\keywords{
Semiconductor device modeling; charge transport models;
hydrodynamic model; upwind discretization; submicron devices; hot electrons;
velocity overshoot, Monte Carlo methods, MESFETs.
}

\section{INTRODUCTION}
There is a growing interest in extended charge transport models for
semiconductor devices. Our paper emerges from the the fact that today's
submicron semiconductor devices
like e.g. MESFETs and HEMTs are operated under strong electric
fields and at high frequencies.
Information transmission using an electromagnetic wave at
very high frequencies will have
a direct impact on how we design active and passive components
in different fields of micro- and optoelectronics.
In such cases, quasi-static semiconductor device models
like the energy balance model (EBM) are no longer adequate.
Especially in GaAs and related materials used for high-speed device design,
inertia effects play an important role since the momentum and energy
relaxation times of the electron gas are close to the picosecond range.

The most elaborate and practicable
approach for the description of charge transport
in semiconductors used for device simulation would be the Monte Carlo (MC)
method \cite{0}. The advantage of this technique is a complete picture of
carrier dynamics with reference to microscopic material parameters,
e.g. effective masses and scattering parameters. But the method must be
still considered as very time consuming and hence too uneconomical
to be used by device designers.

Besides the simplest concept which is the traditional
drift-diffusion model (DDM),
there is a much more rigorous approach to the problem, namely the so-called
hydrodynamic model (HDM). A simplified version of the HDM is the
EBM, which is implemented in most of today's
commercial semiconductor device simulators.
The HDM makes use of electron temperature for the description of charge
transport and takes inertia effects into account as well.
Starting from the Boltzmann equation, Blotekjaer \cite{1}
and many others presented a derivation of such equations
under the assumption of parabolic band structures.
Especially for silicon, satisfactory results were obtained this way
{\cite{5}.
But the results have
often been unsatisfactory when MC models based on a nonparabolic
structure were compared to HDM results based on an empirical choice
of model parameters. Therefore, it is quite natural to improve the
HDM by incorporating energy-dependent relaxation times and effective
masses obtained from MC bulk simulations {\cite{2,3}}. This is a
strategy we pursue in this paper.

In the first part of this work, we give a short definition
of the hydrodynamic model for GaAs.
We emphasize that a thorough analysis of the physical features of
the charge carrier transport models is the basis for a clear understanding
of their limits of applicability.
Then we give a simple discretization scheme for the full hydrodynamic
model in two dimensions, which will be applied to a GaAs MESFET
structure in the last part. There, we compare the HDM results
also to results obtained from the simpler EBM, and we investigate the
influence of the grid resolution on the results. 

\section{THE HYDRODYNAMIC MODEL FOR GaAs FETs}
\subsection{Definition of the model}

Our active device model is based on the {\em{single-gas}}
hydrodynamic equations. This is a simplification of the two-valley
hydrodynamic equations, since strictly speaking, in GaAs and
other semiconductors with similar band structure
like InP, there exists an electron gas
with different thermal distribution function for each conduction band
valley (i.e. the $\Gamma$- and $L$-valleys). The equations for each
valley are, however, coupled through collision terms since electrons can
scatter between two different valleys. The corresponding
relaxation rates may be of
the order of a picosecond and are therefore relatively large. This is the main
drawback of the single-gas approximation, and it would be desirable to
implement at least a two-valley hydrodynamic model. 
Reliable extensive two-valley simulations have been performed only
for the one-dimensional case so far
due to the large amount
of equations and parameters involved in such a model.
A hydrodynamic two-valley simulation of GaAs MESFETs is the subject
of one of our forthcoming papers.
The HDM equations consist of the
continuity equation
\begin{eqnarray}
\frac{\partial n}{\partial t} + \vec{\nabla} (n \vec{v}) = 0 \quad
\label{ce}
\end{eqnarray}
for negligible charge carrier generation and recombination,
the momentum balance equation given by
\begin{equation}
\frac{\partial \vec{p}}{\partial t} + (\vec{\nabla}\vec{p})\vec{v}
+(\vec{p} \, \vec{\nabla})\vec{v}=-en\vec{E} -\vec{\nabla}
(nkT)-\frac{\vec{p}}{\tau_p}
\end{equation}
or alternatively (only for the x-component)
\begin{eqnarray}
\frac{\partial (m^*(\overline{\omega}) n v_x)}{\partial t}+ \vec{\nabla}
(m^*(\overline{\omega}) n v_x \vec{v}) =  \nonumber \\
-q n E_x - \frac{ \partial (n k T)}{\partial x}
-\frac{m^*(\overline{\omega}) n v_x}{\tau_p (\overline{\omega})}
\quad  \label{mb},
\end{eqnarray}
and the energy balance equation
\begin{displaymath}
\frac{\partial \omega}{\partial t}+\vec{\nabla}(\vec{v}\omega) =
\end{displaymath}
\begin{equation}
-en\vec{v}\vec{E} - \vec{\nabla}(nkT\vec{v})-\vec{\nabla}(-\kappa
\vec{\nabla}T) - \frac{\omega-\frac{3}{2}nkT_L}
{\tau_\omega(\overline{\omega})}
\quad ,\label{eb}
\end{equation}
where $n$, $\omega$ ($\overline{\omega}=\omega/n$),
and $\vec{v}$ are the electron density, the electron
energy density (average electron energy) and the electron drift velocity,
respectively. $v_x$ is the x-component of the electron drift velocity
and $\vec{p}=m^*n\vec{v}$ the momentum density.
Corresponding equations are valid for the y- (and z-) components.
$T$ is the electron temperature, $e>0$ the elemental charge
and $k$ Boltzmann's constant.
$\overline{\omega_0}=\frac{3}{2}kT_L$
is the average thermal equilibrium energy of
electrons, where $T_L$ is the lattice temperature.
The EBM uses only a simplified energy balance equation
(see below).
The electronic current density $\vec{J}$ inside the active device
is given by
\begin{eqnarray}
\vec{J}=-e n \vec{v} \quad ,
\end{eqnarray}
the total current density is
\begin{eqnarray}
\vec{J}_{tot}=-e n \vec{v} + \epsilon_0 \epsilon_r
\frac{\partial \vec{E}}{\partial t} \quad .
\end{eqnarray}
The momentum relaxation time $\tau_p(\overline{\omega})$ is related to the mobility of the
electrons via $\mu(\overline{\omega})=(e/m^*(\overline{\omega}))
\tau_p(\overline{\omega})$,
and the energy relaxation time $\tau_\omega(\overline{\omega})$ describes the
exchange of energy between the heated electron gas and the lattice.
$\tau_p$ and $\tau_\omega$ and the effective
electron mass $m^*$ are assumed to be functions of the mean electron
energy. We performed steady-state Monte Carlo simulations in order to
get the correct values for these parameters.

The hydrodynamic equations, together with Poisson's equation
($N_d$ $(\simeq N_d^+)$ is the number of (ionized) donors)
\begin{eqnarray}
\Delta \phi= -\vec{\nabla}\vec{E}=
-\frac{e}{\epsilon_0 \epsilon_r}(N_d^+-n)
\end{eqnarray}
form a complete set of equations that can be used to solve for the
electron density, velocity, energy and electric field
for given boundary conditions,
if we use a closing relation for the mean electron energy
$\overline{\omega}$, the electron temperature $T$ and velocity $v$:
\begin{eqnarray}
\overline{\omega}=\frac{1}{2}m^{**}(\overline{\omega}) v^2 +
\frac{3}{2}k T+\beta_L(\overline{\omega}) \Delta E_{\Gamma L}
\quad  \label{closrel}.
\end{eqnarray}
The reason for the double index of the electron mass will soon become clear.
The last term in eq. (\ref{closrel}) accounts for the fact that
a minimum energy $\Delta E_{\Gamma L}=0.29$ eV is necessary
to excite an electron to an upper conduction band. $\beta_L$ is the
relative fraction of electrons in the L-band for the stationary
homogeneous case. The term $\beta_L(\overline{\omega}) \Delta E_{\Gamma L}$
is often neglected in the literature, but this may lead to an
overestimation of the electron temperature of
{\em{more than 1000 K}} at
high energies.

\subsection{Remarks on the single gas approximation}
We point out again the important fact that we are using a single-gas
approximation for the hydrodynamic model. This means that the closing relation
eq. (\ref{closrel}) is a crude approximation which allows the calculation
of the electron temperature from the total electron energy and electron
drift velocity. Some authors neglect also the influence of the electron
velocity on the temperature {\cite{3,11}} by directly relating the electron
temperature $T$ to the average electron energy
$\overline{\omega}$ from
stationary MC simulations.

The transition from the two-gas model to the single-gas approximation
has to be done carefully;
therefore, we present here a short discussion of the problem
which is usually not mentioned in the literature.
We assume parabolic $\Gamma-$ and $L-$valleys for the sake of brevity,
but in our simulations, we took the effect of the non-parabolicity
of the energy bands into account
by using the Kane model {\cite{12}}, which generalizes
the parabolic relation for the electron energy $E_k$ and electron crystal
momentum $\hbar k$
\begin{eqnarray}
E_k=\frac{\hbar^2 k^2}{2 m_{\mbox{\small{eff}}}}=\gamma(k)
\end{eqnarray}
to
\begin{eqnarray}
E_k(1-\alpha E_k)=\gamma(k) \quad ,
\end{eqnarray}
with the non-parabolicity coefficient $\alpha$ which has different
values for the different energy bands. $m_{\mbox{\small{eff}}}$ is the
effective electron mass at the bottom of the energy band under
consideration. Very often, an energy-dependent electron mass
is defined via
\begin{eqnarray}
\frac{1}{m_{\mbox{{\small{E}}}}} &=&\frac{1}{\hbar^2}
\frac{\partial^2 E_k}{\partial k^2}\\
&=& \frac{1}{m_{\mbox{\small{eff}}}} \frac{1}{(1+4 \alpha
\gamma(k))^{3/2}} \quad ,
\end{eqnarray}
i.e. the electron mass increases with growing energy.
The electron mass
$m_E$ must not be confused with the energy-dependent electron mass
used in the hydrodynamic model. In that case, the masses are a kind of
{\em{average}} masses depending on the {\em{average}} electron energy.

There is even a further aspect related to the notion of the electron mass.
The crystal velocity of an electron is given for spherical bands
by
\begin{eqnarray}
v=\frac{1}{\hbar} \frac{\partial E_k}{\partial k} \quad .
\end{eqnarray}
A short calculation for the Kane model shows that
\begin{eqnarray}
v=\frac{\hbar k}{m_{\mbox{\small{eff}}}} \frac{1}{\sqrt{1+4\alpha
\gamma(k)}} \quad ,
\end{eqnarray}
which implies that crystal velocity $v$ and crystal momentum
$p=\hbar k$ are related by
\begin{eqnarray}
p=m_p v \quad, \quad m_p=\sqrt{1+4\alpha \gamma(k)}
m_{\mbox{{\small{eff}}}} \quad ,
\end{eqnarray}
i.e. a different definition of the energy-dependent electron mass
applies if the electron velocity is calculated from the crystal
momentum.

In the single particle two-band MC simulations, the random walk
of an electron inside the semiconductor material is monitored
over a sufficiently long time. As a result,  
the probability $\beta_\Gamma$ that an electron resides in the
$\Gamma-$valley is obtained as a function of the applied constant
homogeneous electric
field $E$ or as a function of the mean electron energy $\overline{\omega}$,
and the probability
of finding the electron in an upper $L-$valley is
then $\beta_L=1-\beta_\Gamma$.
Also the values for
for the average electron velocities $v_\Gamma$ and $v_L$ 
in the different valleys are obtained as well.
Then it is reasonable to define the average electron velocity
by
\begin{eqnarray}
v=\beta_\Gamma v_\Gamma + \beta_L v_L \quad .
\end{eqnarray}
The average electron momentum $p$ is given by
\begin{eqnarray}
p=m^* v =m_\Gamma \beta_\Gamma v_\Gamma +m_L \beta_L v_L
\quad ,
\end{eqnarray}
hence the (energy-dependent) electron mass which must be used
in the hydrodynamic model in order to relate average electron
velocity and electron momentum is calculated from
\begin{eqnarray}
m^*=\frac{m_\Gamma \beta_\Gamma v_\Gamma +m_L \beta_L v_L}
{\beta_\Gamma v_\Gamma+\beta_L v_L} \quad .
\end{eqnarray}
But we suggest that a {\em{different mass}} $m^{**}(\overline{\omega})$
should be used for the
calculation of the average kinetic electron energy in eq. (\ref{closrel}).
We identify
\begin{eqnarray}
\frac{1}{2}m^{**}v^2=\frac{1}{2}\beta_\Gamma m_\Gamma v_\Gamma^2+
\frac{1}{2} \beta_L m_L v_L^2 \quad ,
\end{eqnarray}
and therefore
\begin{eqnarray}
m^{**}=\frac{m_\Gamma \beta_\Gamma v_\Gamma^2 +m_L \beta_L v_L^2}
{(\beta_\Gamma v_\Gamma+\beta_L v_L)^2}
\quad .
\end{eqnarray}
$m^*$ and $m^{**}$ are {\em{not distinguished}} in the literature.
It is tempting to use a naive definition for the electron mass
\begin{eqnarray}
m=\beta_\Gamma m_\Gamma+\beta_L m_L \quad.
\end{eqnarray}
It is interesting that data in the literature
for the energy-dependent electron mass
are usually in better agreement with this definition.
It is clear that the HDM is still an approximative description
of charge carrier dynamics inside a semiconductor, and the different
assumptions which are inherent in the derivation of the model may already
cause larger errors in the simulation results than using only one mass.
Therefore we do not claim that our discussion leads to improved
simulation results, but it rather shows the limits of the frequently
used model and points out that it is mandatory to maintain
always a highest possible degree of consistency.

Figure 1 shows $m^*$, $m^{**}$ and $m$ normalized to the free
electron mass $m_0$ as functions of the average electron energy
for a GaAs lattice temperature $T_L=300$ K and a low
doping density $N_d=10^{14} \mbox{cm}^{-3}$.
The results from MC simulations were smoothed by a polynomial fit
and transferred afterwards into the hydrodynamic simulation program.
We used $m_\Gamma=0.067 m_0$ and $m_L=0.35 m_0$. For high energies
when nearly all electrons are in the upper bands, the electron mass
even exceeds the value $m_L$ due to the non-parabolicity of
the energy bands.

\begin{figure}
\centering\includegraphics[width=10 cm]{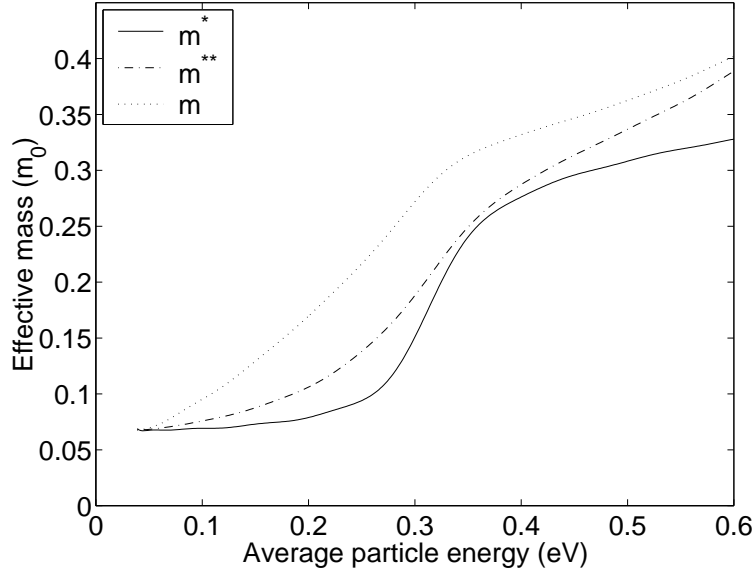}
\caption{Energy-dependent electron masses used in the hydrodynamic
simulations.}
\end{figure}

Figure 2 shows the average energy $\overline{\omega}$
of an electron in a constant homogeneous
electric field E for GaAs. For each data point, the electron was
scattered one million times (including so-called self-scattering),
therefore the resulting curve is already quite smooth.
It is clear that MC simulations deliver no data for average
electron energies below $\overline{\omega}<\overline{\omega_0}=38$ meV,
since the mean electron
energy $\overline{\omega}$ has this value if no electric field is
applied to the crystal, and $\overline{\omega}$ grows for increasing
electric field. This is no major problem, since the low energy region
is of minor importance for the hydrodynamic simulation and the
necessary data can be obtained from theoretical considerations {\cite{2}}.

\begin{figure}
\centering\includegraphics[width=10 cm]{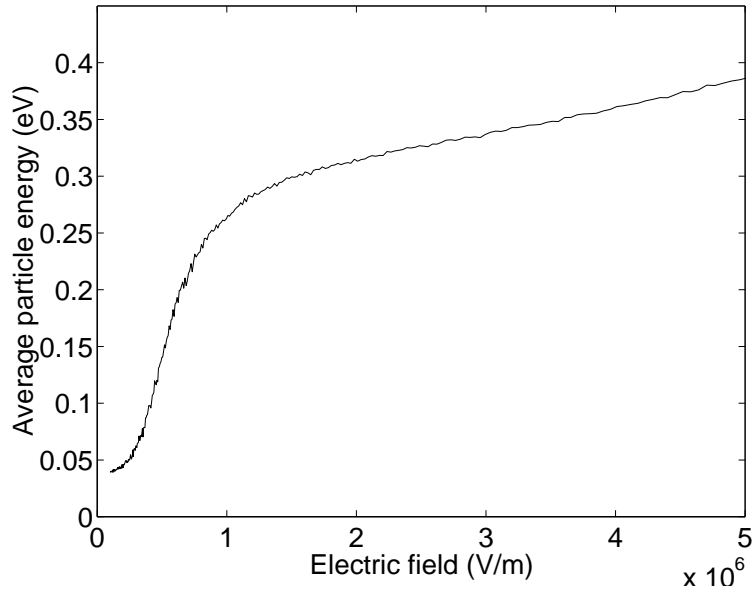}
\caption{E-$\overline{\omega}$-relation for GaAs with $T_L=300$ K
and $N_d=2 \cdot 10^{17} \mbox{cm}^{-3}$.}
\end{figure}

In order to complete the set of data which is necessary for
hydrodynamic simulations, the electron velocity and energy relaxation
times are depicted in Figs. 3 and 4 for doping densities
$N_d=10^{14} \mbox{cm}^{-3}$ and $2 \cdot 10^{17} \mbox{cm}^{-3}$.
The characteristic shape of the velocity curve can be explained
by the fact that at high energies the electrons jump into the $L-$bands
where the electrons have a lower mobility than in the $\Gamma-$band.

\begin{figure}
\centering\includegraphics[width=10 cm]{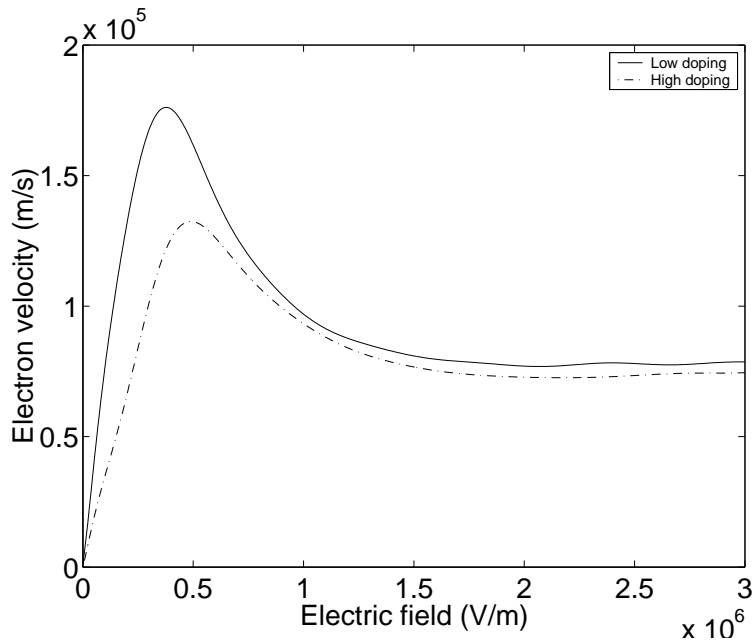}
\caption{v-E curve for GaAs.}
\end{figure}
\begin{figure}
\centering\includegraphics[width=10 cm]{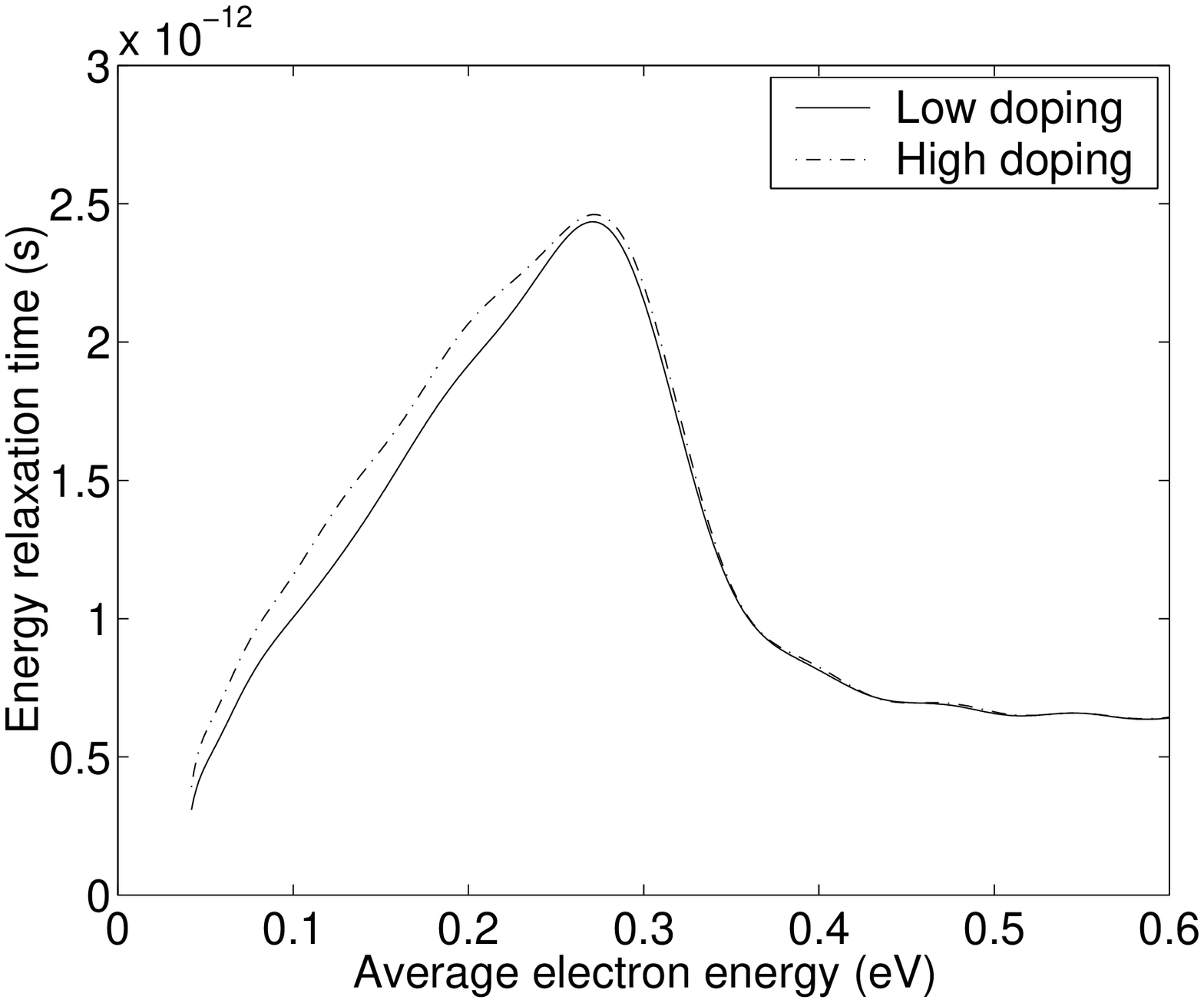}
\caption{Energy relaxation times for GaAs (from a MC simulation,
as in Figs. 1-3)}
\end{figure}

Finally, we need an expression for the thermal conductivity of
the electron gas, which is given by theoretical considerations
\begin{eqnarray}
\kappa=(5/2+r)n \frac{k^2 \mu(\overline{\omega})}{e} T \quad . \label{heat}
\end{eqnarray}
Several different choices for $r$ can be found in the literature,
and many authors
{\cite{2,3}} even neglect heat conduction in their models.
As a matter of fact, heat conduction does not influence the
simulation results very much if $r$ remains within a reasonable range,
but Baccarani and Wordeman point out in {\cite{4}} that neglecting
thermal conductivity completely
can lead to nonphysical results and mathematical instability.
Although their work is directed to Si, their remarks should be equally
valid for GaAs since the equations have a similar form in both cases.
In our simulations, we have chosen $r=-2$.

\subsection{The energy balance model}
The EBM is obtained as a simplification of the full HDM
by neglecting the convective terms of the momentum balance equation
(\ref{mb}).
Additionally, the energy balance equation (\ref{eb}}) is simplified by the
assumption that the time derivative of the mean electron energy
$\partial \omega/\partial t$ is small compared to the other terms and that
the kinetic part in $\omega$ can also be neglected, i.e.
\begin{equation}
\omega=\frac{3}{2}nk T \quad,
\end{equation}
or, if we take the two-valley structure of GaAs into account,
\begin{eqnarray}
\overline{\omega}=
\frac{3}{2}k T+\beta_L(\overline{\omega}) \Delta E_{\Gamma L}
\quad.
\end{eqnarray}
The energy balance equation then becomes
\begin{displaymath}
\vec{\nabla}(\vec{v}\omega) =
\end{displaymath}
\begin{equation}
-en\vec{v}\vec{E} - \vec{\nabla}(nkT\vec{v})-\vec{\nabla}(-\kappa
\vec{\nabla}T) - \frac{\omega-\frac{3}{2}nkT_L}
{\tau_\omega(\overline{\omega})}
\end{equation}
and the momentum balance equation simplifies to the well-known current
equation
\begin{equation}
\vec{j}=-\frac{e}{m}\tau_p n \vec{E}-\frac{e}{m} \tau_p \vec{\nabla}
\Bigl( \frac{nkT}{e} \Bigr) \quad .
\end{equation}
The continuity equation and the Poisson equation remain of course
unchanged in the EBM.
Neglecting the time derivative of the current density is equivalent
to the assumption that the electron momentum is able to adjust itself
to a change in the electric field within a very short time. While this
assumption is justified for relatively long-gated field effect
transistors, it needs to be investigated for short-gate cases.

Setting the electron temperature $T$ equal to the (constant) temperature
of the semiconductor material $T_L$ leads to the drift-diffusion model.
Such a simplification is clearly not justified for the case studied
in this paper (see also \cite{aste}).

\section{NUMERICAL ASPECTS}
\subsection{Discretization of the equations}

Today, many elaborate discretization methods are available for
the DDM equations or EBM equations. The well-known Schar\-fet\-ter-Gum\-mel
method \cite{9} for the DDM makes use of the fact that the current
density is a slowly varying quantity. The current equation is then
solved exactly under the assumption of a constant current density over
a discretization cell,
which leads to an improved expression for the current density than it is
given by simple central differences. It is therefore possible to
implement physical arguments into the discretization method.
Similar techniques have been worked out for the EBM \cite{10}.
But due to the complexity of the HDM equations, no
satisfactory discretization methods which include physical input
are available for this case.

Therefore, we developed a shock-capturing
upwind discretization method, which has the
advantage of being simple and reliable.
For our purposes, it was sufficient to
use a homogeneous mesh and a constant time step.
But the method can be generalized to the non-homogeneous case.

Time discretization is done for
all equations by forward Euler differencing, i.e. the discretization
scheme is fully explicit. The discretization is always written down only for
the x-component of vectorial quantities in the sequel,
since the corresponding
expressions for y-components are then easy to derive.

The constant timestep $\Delta t$ used in our
simulations was typically of the order of a
few tenths of a femtosecond, and quantities at time $T=t \Delta t$ carry
an upper integer index $t$.

The rectangular simulated region of the MESFET gets discretized into
$N_x \times N_y$ rectangular cells $C_{i,j}$ of equal size $\Delta x
\times \Delta y=(l_x/N_x) \times (l_y/N_y)$.
Scalar quantities 
at timestep $t$ like
$n^t_{i,j},\omega^t_{i,j},T^t_{i,j}$ and $\phi^t_{i,j}$,
where $i=1,...N_x$ and $j=1,...N_y$
are located at the center of the cells, whereas
vectorial quantities like e.g. the electric field components
$E^t_{x;i+1/2,j}$, $E^t_{y;i,j+1/2}$ or the velocity components
$v^t_{x;i+1/2,j}$, $v^t_{y;i,j+1/2}$ are always calculated first at
midpoints between the scalar quantities.

If necessary,
we can define intermediate values, e.g. $E_{x;i,j}$ by
\begin{equation}
E_{x;i,j}=\frac{1}{2}(E_{x;i-1/2,j}+E_{x;i+1/2,j}) \quad ,
\end{equation}
but a different definition applies to e.g. $j_{x;i,j}$,
as we shall see.

The fundamental variables that we will have to compute at each timestep
are $n_{i,j}$, $\phi_{i,j}$, $\omega_{i,j}$ (or $T_{i,j}$), $j_{x;i+1/2,j}$
and $j_{y;i,j+1/2}$,
always respecting the imposed boundary conditions. All other variables
used in the sequel should be considered as derived quantities.

The momentum balance equation is discretized
in the following way:
\begin{eqnarray}
\frac{p^{t+1}_{x;i+1/2,j}-p^t_{x;i+1/2,j}}{\Delta t} =
-qn^t_{i+1/2,j}E^t_{x;i+1/2,j} \nonumber \\
-\frac{k}{\Delta x}(n^t_{i+1,j}T^t_{i+1,j}-
n^t_{i,j}T^t_{i,j})/n^t_{i+1/2,j} \nonumber \\
-(p_{x;i+1/2,j}^t v_{x;i+1/2,j}-p_{x;i-1/2,j}^t v_{x;i-1/2,j})/\Delta x\\
-(p_{x;i+1/2,j}^t v_{y;i,j+1/2}-p_{x;i+1/2,j-1}^t v_{y;i,j-1/2})/\Delta y\\
-p^t_{x;i+1/2,j}/\tau^t_{p;i+1/2,j} \quad ,
\end{eqnarray}
where $p_{x;i+1/2,j} > 0$ and $p_{y;i,j+1/2} > 0$
and the same discretization strategy is applied to the y-component of
the electron velocity. From the momentum density we obtain the
new particle current density by
\begin{eqnarray}
j_{x;i+1/2,j}^{t+1}=p_{x;i+1/2,j}^{t+1}/m^{*t}_{i+1/2,j} \quad ,
\end{eqnarray}
and
the momentum density at $(i,j)$ is extrapolated from neighbouring points
in the direction of the electron flow x-component
\begin{equation}
p_{x;i,j}^{t+1}=\left\{ \begin{array}{r@{\quad:\quad}l}
\frac{3}{2} p_{x;i-1/2,j}^{t+1}-\frac{1}{2} p_{x;i-3/2,j}^{t+1}) &
p_{x;i+1/2,j}^{t+1} >0 \\
\frac{3}{2} p_{x;i+1/2,j}^{t+1}-\frac{1}{2} p_{x;i+3/2,j}^{t+1}) &
p_{x;i+1/2,j}^{t+1} < 0
\end{array} \right. \quad ,
\end{equation}
and finally we get
\begin{equation}
v_{x;i,j}^{t+1}=p_{x;i,j}^{t+1}/n_{i,j}^t/m^{*t}_{i,j} \quad ,
\end{equation}
\begin{equation}
v_{x;i+1/2,j}^{t+1}=j_{x;i+1/2,j}^{t+1}/n_{i+1/2,j}^t/m^{*t}_{i+1/2,j}\quad .
\end{equation}
We found that the purely heuristic choice
\begin{equation}
n^t_{i+1/2,j}=\sqrt{n^t_{i,j} n^t_{i+1,j}}
\end{equation}
in the equations above improves the stability of our code.

The electron temperature is related to the energy density
by the relation $\omega_{i,j}^t=\frac{3}{2}n^t_{i,j}k T_{i,j}^t+
\frac{1}{2}m^{**t}_{i,j}n_{i,j}^t (v_{x;i,j}^{t2}+v_{y;i,j}^{t2})
+\beta^t_{L;i,j} \Delta E_{\Gamma L}$
and can therefore be regarded as a dependent variable.
The upwind discretization is of the energy balance equation is given by
\begin{eqnarray}
\frac{\omega_{i,j}^{t+1}-\omega_{i,j}^t}{\Delta t} & = &
-en_{i,j}^t(v_{x;i,j}^{t+1}E_{x;i,j}^t
+v_{y;i,j}^{t+1}E_{y;i,j}^t)  \nonumber \\
& & -\frac{\omega_{i,j}^t-\frac{3}{2}n_{i,j}^tkT_L}{\tau_{\omega;{i,j}}^t}
\nonumber \\
& & -\frac{1}{\Delta x}(j_{x;e,i+1/2,j}^t-j_{x;e,i-1/2,j}^t) \nonumber \\
& & -\frac{1}{\Delta x}(j_{x;p,i+1/2,j}^t-j_{x;p,i-1/2,j}^t) \nonumber \\
& & -\frac{1}{\Delta x}(j_{x;h,i+1/2,j}^t-j_{x;h,i-1/2,j}^t) \nonumber \\
& & -\frac{1}{\Delta y}(j_{y;e,i,j+1/2}^t-j_{y;e,i,j-1/2}^t) \nonumber \\
& & -\frac{1}{\Delta y}(j_{y;p,i,j+1/2}^t-j_{y;p,i,j-1/2}^t) \nonumber \\
& & -\frac{1}{\Delta y}(j_{y;h,i,j+1/2}^t-j_{y;h,i,j-1/2}^t) \quad ,
\end{eqnarray}
where we have defined the energy currents
\begin{equation}
j_{x;e,i+1/2,j}^t=v_{x;i+1/2,j}^{t+1} \omega_{i+1/2,j}^t \quad ,
\end{equation}
\begin{equation}
j_{x;p,i+1/2,j}^t=kj_{x;i+1/2,j}^{t+1} T_{i+1/2,j}^t \quad ,
\end{equation}
and
\begin{equation}
j_{x;h,i+1/2,j}^t=-\kappa_{i+1/2,j}^t (T_{i+1,j}^t-T_{i,j}^t)/\Delta x \quad .
\end{equation}
Having obtained the new values for the mean electron energy,
the transport parameters and energy-dependent masses are then also updated.

The current continuity equation is discretized in a conservative way,
using the particle current density $\vec{j}=n\vec{v}$
\begin{displaymath}
\frac{n^{t+1}_{i,j}-n^t_{i,j}}{\Delta t} = -(j^{t}_{x;i+1/2,j}-
j^{t}_{x;i-1/2,j})/\Delta x
\end{displaymath}
\begin{eqnarray}
 - (j^{t}_{y;i,j+1/2}-j^{t}_{y;i,j-1/2})/
\Delta y \quad , \label{update}
\end{eqnarray}
i.e. particles that leave cell $(i,j)$ in x-direction enter cell $(i+1,j)$
and analogously for the y-direction; therefore, the total number of
electrons inside the MESFET can only be changed at the boundary of
the simulation region (mainly at the contacts).

The Poisson equation, which is coupled to the hydrodynamic equations
only through the particle density $n$, can be solved by any convenient
method which is fast enough, since the computational effort should
be kept as small as possible.
Therefore, we used a multigrid method to perform this task. Fortunately, the Poisson equation has not to be solved
at each timestep. Since the relaxation times of GaAs are of the order
of some tenths of a picosecond, a timestep of about ten femtoseconds is fully
sufficient for the update of the electric field. Fortunately,
the stability of the discretization scheme is not affected that way, and allows
an enormous reduction of the computational costs. 

\subsection{Boundary conditions}
For the basic quantities $n,\phi$ and $\omega$ we imposed
Dirichlet boundary conditions at each timestep at the contacts, e.g.
the potential $\phi$ at the source and the drain was fixed
by the applied voltage
\begin{equation}
\phi|_{s,d}=V_{s,d}.
\end{equation}
Analogously, we assumed charge neutral contacts at the
source and the drain, such that the charge carrier density
was given there by the fixed doping density.
The gate contact was modelled
by assuming a Schottky barrier height of 0.8 V. For further
details concerning the metal-semiconductor contact modelling
we refer to standard textbooks \cite{0}.
As far as the energy density is concerned, we imposed the
boundary conditions directly on the electron temperature by assuming
that the electron gas is in thermal equilibrium with the
drain/source contacts.

The artificial boundaries were modelled using von Neumann type
boundary conditions.
We present an explicit example in order to illustrate this point.
We assume that the discrete values of the electron density
at the MESFET boundary between
the source and the gate (see Fig. 5) are given by
$n_{(i_1,1)},...n_{(i_2,1)}$:
The first index denotes the horizontal
direction, whereas the second index starts with $1$
at the top of the MESFET.
Then, after each update of the density
according to eq. (\ref{update}), we enforce
\begin{equation}
n_{(i_1,1)},...n_{(i_2,1)}=n_{(i_1,2)},...n_{(i_2,2)},
\end{equation}
corresponding to the von Neumann condition that the  
normal component of the electron density vanishes at the
specified boundaries.

As mentioned above, we used a multigrid algorithm to calculate
the electric potential. Also there, the mixed  Dirichlet/von Neumann
boundary conditions were imposed on the subgrids at each intermediate
step of the calculations. A FORTRAN90 program which calculates the
potential by a multigrid algorithm can be obtained from the first
author's address.

\section{SIMULATION RESULTS}
The GaAs MESFET structure used in our simulation is shown in Fig. 5.
The structure consists of a 0.1 $\mu$m-thick active layer with a
doping concentration of $N_d=2 \cdot 10^{17} \mbox{cm}^{-3}$ on a
0.3 $\mu$m buffer layer ($N_d=10^{14} \mbox{cm}^{-3}$). The doping profile
is abrupt between the two layers, the lattice temperature is $T_L=300$ K.
For steady-state results, we used
long simulation times of 30 ps such that the steady state was {\em{de facto}}
reached. The length of the drain and source contacts is 0.5 $\mu$m,
the gate-source separation 0.5 $\mu$m, the gate-drain separation is 1.0
$\mu$m and the gate length is 0.8 $\mu$m. The Schottky barrier height is
assumed to be 0.8 V.

\begin{figure}
\centering\includegraphics[width=10 cm]{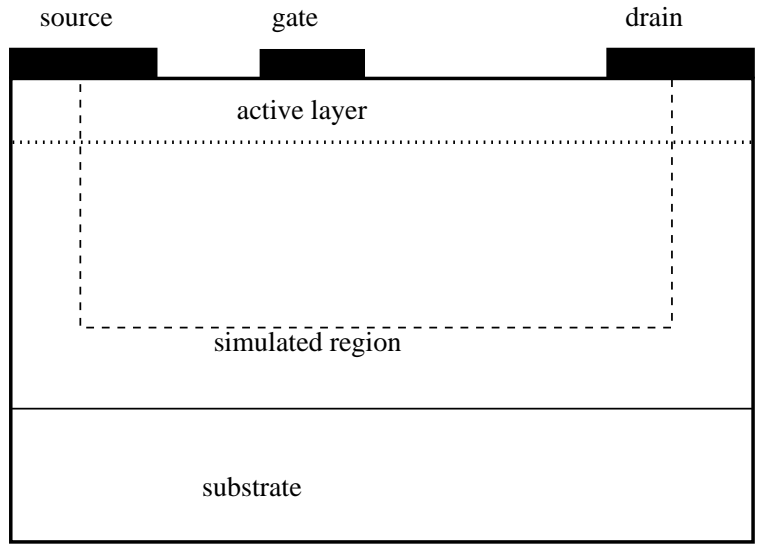}
\caption{The MESFET geometry.}
\end{figure}

In order to obtain stable and physically meaningful results, values
like $\Delta x=\Delta y=6.1$ nm for a grid size of
$537 \times 65$ were typical values used in the simulations. The rather
large mesh was manageable due to the effective multigrid algorithm
used for the solution of the Poisson equation. We found that the
mesh size used in {\cite{3}} was too coarse for accurate results,
although the authors improve accuracy and convergence speed
of their calculations by using a non-homogeneous grid. In fact,
a non-homogeneous grid necessitates additional calculational costs which
reduce the speed of the simulation, and the timesteps also depend on the
size of the smallest cells.
Furthermore, the Poisson equation was solved
by a conventional successive overrelaxation method.

We present results for a gate-source bias $V_{gs}=0$ V and
a drain-source bias $V_{ds}=3$ V.

Fig. 6 shows the electron temperature inside the MESFET for the stationary
case.

\begin{figure}
\centering\includegraphics[width=10 cm]{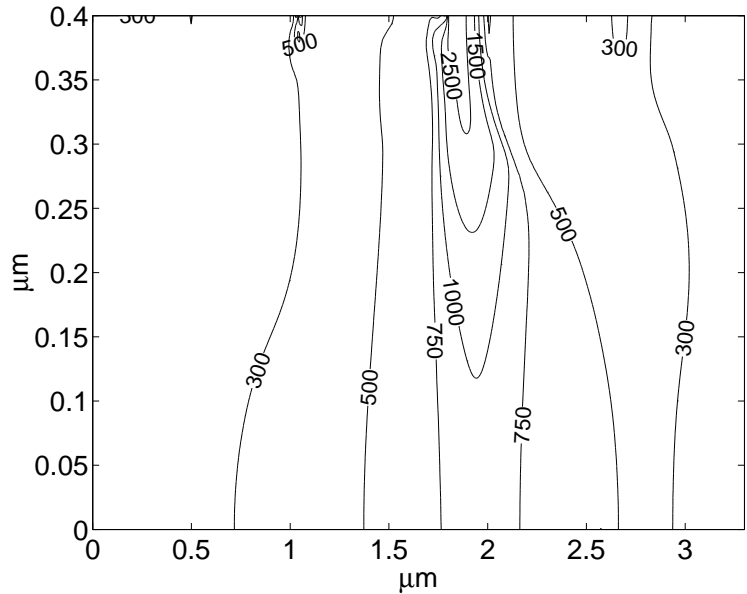}
\caption{Electron temperature inside the MESFET.}
\end{figure}

Fig. 7 and 8 show the electron velocity and the electron temperature
along the channel of the MESFET (0.077 $\mu$m below the contacts).
Due to the high temperature which
is reached under the gate, the electron mobility is strongly lowered
in this region. The electrons, which overshoot under the gate, are
therefore deaccelerated abruptly to a lower velocity in the 
high-temperature region. The results obtained from an energy
balance calculation are in good quantitative agreement, the
differences are mostly pronounced in the region where the electron
velocity is high, as it is expected from the different
treatment of the energy density in the HDM and EBM.

In the very close vicinity of the gate, it was necessary to reduce
artificially the electric field or the velocity of the electrons
in order to stabilize our code. We checked that this does not 
strongly affect the simulations results outside this region
due to the very low density of the electron gas near the contact;
a similar procedure was also necessary in {\cite{3,11}}.

The velocity and temperature curves are in fact very similar to those of
one-dimensional simulations of ballistic diodes ($n^+-n-n^+$-structures),
which were used as simplified models for FET channels.

\begin{figure}
\centering\includegraphics[width=10 cm]{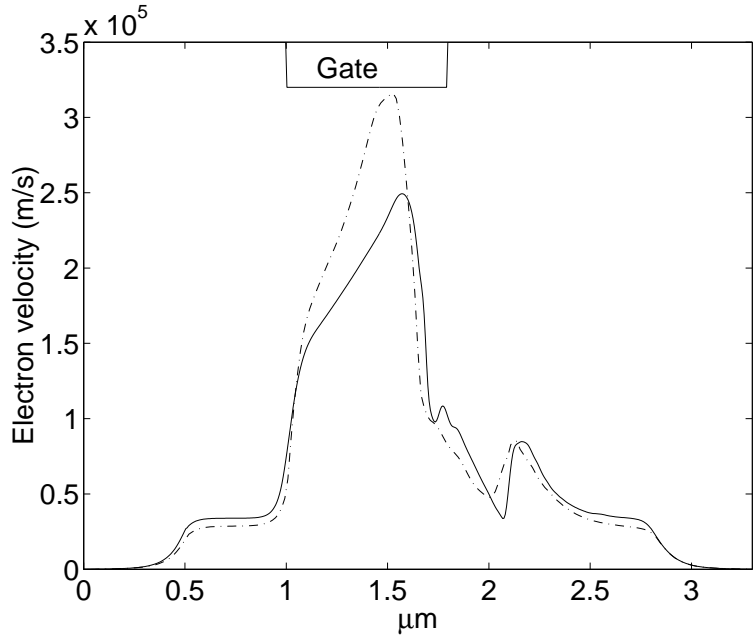}
\caption{Electron velocity inside the MESFET for a fine grid
(solid line) and a coarse grid (dash-dotted).}
\end{figure}

\begin{figure}
\centering\includegraphics[width=10 cm]{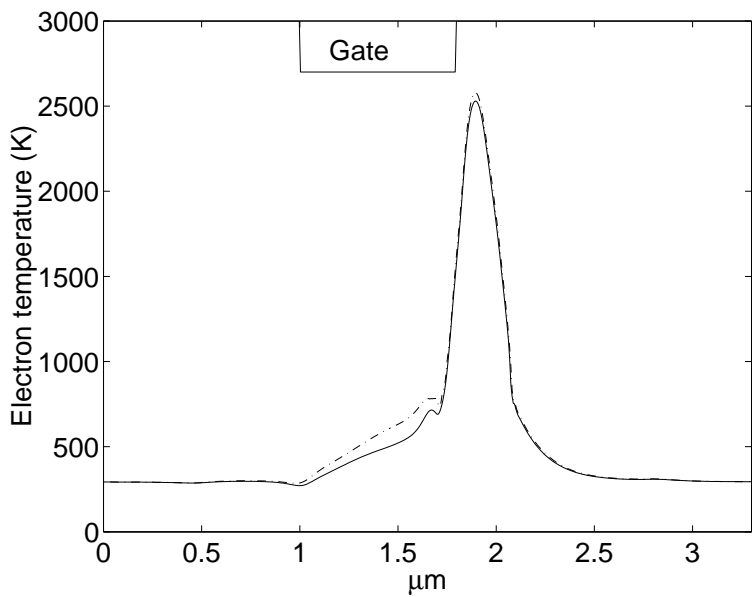}
\caption{Electron temperature inside the MESFET along
the channel. The dash-dotted line shows the EBM result.}
\end{figure}

It is interesting to observe in Fig. 7 that our simulation results for
a coarse grid (grid size $217 \times 33$) are very close to
those presented in Fig. 5 in \cite{3}, where a non-uniform
mesh of typical size $141 \times 35$ was used for a similar
MESFET geometry. This may be considered as a confirmation of
the results given in \cite{3} for a relatively coars grid,
but demonstrates the fact that stability
of the code does not automatically imply accuracy of the
results, and an investigation of the dependence of the
results on the grid resolution is indispensable.
The channel below the gate (see also Fig. 10), which is the most
interesting region in the MESFET, is relatively small and
must be resolved sufficiently.

Due to the strong heating of the electron gas in the channel region,
most electrons are excited to the L-band. This leads to a cooling effect
of the electron gas, since the excitation energy is missing in the
thermal energy balance.
The dash-dotted curve in
Fig. 9 shows the electron temperature that would be obtained
from the HDM if the energy term $\beta_L(\overline{\omega}) 
\Delta E_{\Gamma L}$ in
eq. (\ref{closrel}) were neglected. The energy
$\Delta E_{\Gamma L}=0.29$ eV which is necessary
to excite an electron to the upper conduction band corresponds
to a temperature difference
$\Delta T=2\Delta E_{\Gamma L}/3k \sim 2000 K$;
the observed error is of the same size.

\begin{figure}
\centering\includegraphics[width=10 cm]{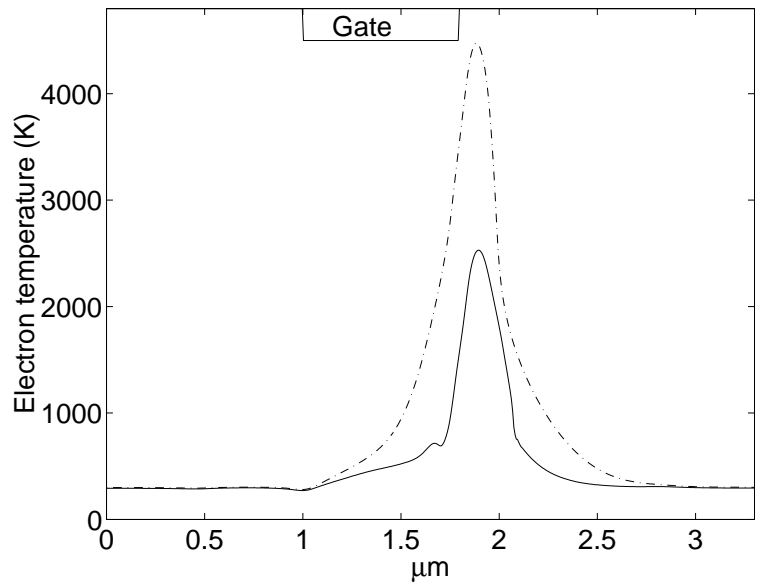}
\caption{Electron temperature inside the MESFET along
the channel for the 'correct' model (solid line) and
the 'wrong' model where the excitation of electrons
to the upper conduction band is not taken into account properly.}
\end{figure}

Finally, Fig. 10 shows a surface plot of the electron density inside the
device. Clearly visible is the MESFET channel under the gate.

\begin{figure}
\centering\includegraphics[width=10 cm]{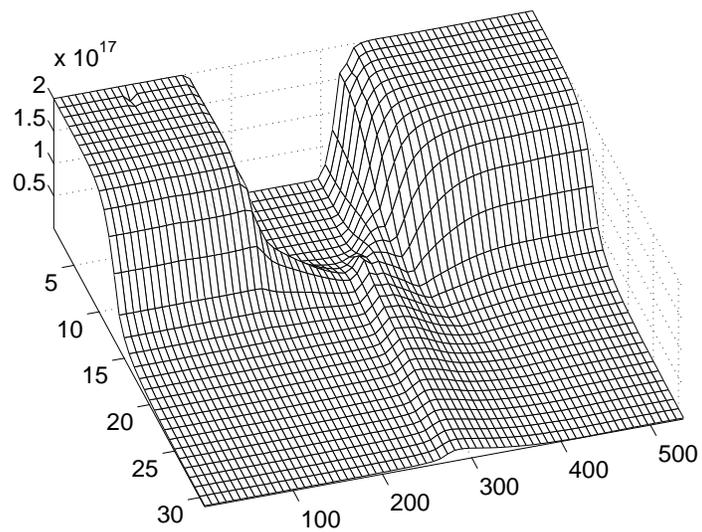}
\caption{Electron density (electrons per $\mbox{cm}^{3}$).
The plot extends over the whole length of the device (537
grid points), but shows only a relevant layer of about
0.2 $\mu$m below the contacts.}
\end{figure}

\section{CONCLUSION}
The feasibility of two-dimensional hydrodynamic simulations is
demonstrated for the case of a GaAs MESFET structure. 
Although the single-gas hydrodynamic model is superior to the
drift-diffusion or energy balance model, it is desirable to direct
the efforts of future research in the direction of multi-valley
hydrodynamic models. Models like the EBM will no longer
be adequate for the physical description of high-speed submicron
devices in the near future.

It is obvious that future attempts to
model submicron devices will face many more problems which
have not been touched in this paper.
One difficulty is the fact that
the components of semiconductor devices are often of very
different size and material composition.
This necessitates the use of adaptive discretization grids or
the hybridization of different numerical methods, but both strategies
are hampered by severe problems like numerical instabilities
or huge computational efforts for realistic simulations.
Another class of problems is due to the fact that the physics
of semiconductor materials is very complex,
and therefore hard to implement such that the physical
behavior of the device is satisfactorily described.
Attempts to describe quantum effects in device modelling should
be considered as tentative in the case of heterostructure
devices.
There is no optimal solution for these problems,
and the numerical and physical models
have to be adapted to the problem under study. We hope that our
detailed description of a hydrodynamic simulation may serve also
as a help for researchers entering the field of hydrodynamic
semiconductor device modelling.

%

%

\subsection*{Biographies}

{\bf Andreas Aste} received the diploma degree in theoretical
physics from the University of Basel, Basel, Switzerland, in 1993,
and the Ph.D. degree
from the University of Z\"urich, Z\"urich, Switzerland, in 1997.
From 1997 to 1998 he was a post doctoral assistant at the
Institute for Theoretical
Physics in Z\"urich. From 1998 to 2001
he was a research assistant and Project
Leader in the Laboratory for Electromagnetic Fields and Microwave
Electronics of the Swiss Federal Institute of Technology ETH.
Since 2001, he is working as a researcher at the Institute for
Theoretical Physics at the University of Basel.
Dr. Aste is a member of the American Physical Society APS.

\vskip 0.4 cm

{\bf R\"udiger Vahldieck} received the Dipl.-Ing. and Dr.-Ing.
degrees in
electrical engineering from the University of Bremen, Germany,
in 1980 and 1983, respectively.
From 1984 to 1986, he was a Research Associate at the University of Ottawa,
Ottawa, Canada. In 1986, he joined the Department of Electrical and
Computer Engineering, University of Victoria,
British Columbia, Canada, where he became a
Full Professor in 1991. During Fall and Spring 1992-1993, he was visiting
scientist at the Ferdinand-Braun-Institute f\"ur Hochfrequenztechnik in
Berlin, Germany. Since 1997, he is Professor of field theory at the
Swiss Federal Institute of Technology, Z\"urich, Switzerland.
His research interests include numerical methods to model electromagnetic
fields in the general area of electromagnetic compatibility (EMC) and
in particular for computer-aided design of microwave, millimeter wave
and opto-electronic integrated circuits.

Prof. Vahldieck, together
with three co-authors, received the 1983 Outstanding Publication Award
presented by the Institution of Electronic and Radio Engineers. In 1996,
he received the 1995 J. K. Mitra Award of the Institution of Electronics
and Telecommunication Engineers (IETE) for the best research paper.
Since 1981 he has published more
than 230 technical papers in books, journals and conferences,
mainly in the field of microwave CAD. He is the Past-President
of the IEEE 2000 International Zurich Seminar on Broadband
Communications (IZS'2000) and since 2003 President and
General Chairman of the international Zurich Symposium on
Electromagnetic Compatibility. He is a member of the
editorial board of the IEEE Transaction on Microwave Theory
and Techniques. From 2000 until 2003 he served as an
Associate Editor for the IEEE Microwave and Wireless Components
Letters and from January 2004 on as the Editor-in-Chief.
Since 1992 he serves on the Technical Program Committee of
the IEEE International Microwave Symposium, the MTT-S Technical
Committee on Microwave Field Theory, and in 1999 on the TPC of
the European Microwave Conference. From 1998 until 2003
Professor Vahldieck was the chapter chairman of the
IEEE Swiss Joint Chapter on MTT, AP and EMC.

\vskip 0.4 cm
{\bf Marcel Rohner} received the Dipl.-Ing. and Dr.-Ing.
degrees in electrical engineering from the Swiss Federal Institute of
Technology, Switzerland, in 1993 and 2002, respectively.
Since 1993 he was a research assistant in the
Electronics Laboratory of the Swiss Federal Institute of Technology ETH,
where he has been working in the area of semiconductor device modeling,
Monte-Carlo device simulations, electro-optical sampling,
digital filtering, and switched-current circuits.
In 2002, he joined the corporate technology center of Leica
Geosystems AG in the field of electronic distance measurements.


\begin{thebibliography}{1}
\bibitem{0}
K. Tomizawa, {\em{Numerical simulation of submicron semiconductor
devices}}. Artech House: London, Boston, 1993.
\bibitem{1}
K. Blotekjaer,"Transport equations for electrons in two-valley
semiconductors,"
{\em{IEEE Trans. Electron Dev.}}, vol. 17, no. 1, pp. 38-47, Jan. 1970.
\bibitem{5}
C.L. Gardner, "Numerical simulation of a steady-state electron shock wave
in a submicrometer semiconductor device,"
{\em{IEEE Trans. Electron Dev.}}, vol. 38, no. 2, pp. 392-398, Feb. 1991.
\bibitem{2}
Y.K. Feng, A. Hintz, "Simulation of submicrometer GaAs MESFETs using
a full dynamic transport model," {\em{IEEE Trans. Electron Dev.}},
vol. 35, no. 9, pp. 1419-1431, Sept. 1988.
\bibitem{3}
M.A. Alsunaidi, S.M. Hammadi, S.M. El-Ghazaly,
"A parallel implementation
of a two-dimensional hydrodynamic model for microwave semiconductor device
including inertia effects in momentum relaxation,"
{\em{Int. J. Num. Mod.: Netw. Dev. Fields}}, vol. 10, no. 2, pp. 107-119,
March-April 1997.
\bibitem{11}
S.M. El-Ghazaly, {\em{private communication}}.
\bibitem{12}
E.O. Kane, "Band structure of Indium Antimonide,"
{\em{J. Phys. Chem. Solids}}, vol. 1, pp. 249-261, 1957.
\bibitem{4}
G. Baccarani, M.R. Wordemann, "An investigation of steady-state
velocity overshoot in silicon,"
{\em{Solid-State Electronics}}, vol. 28, pp. 407-416, 1985.
\bibitem{aste}
A. Aste, R. Vahldieck, "Time-domain simulation of the full hydrodynamic
model",{\em{Int. J. Num. Mod.: Netw. Dev. Fields}},
vol. 16, no.2, pp. 161-174, 2003.
\bibitem{9}
D.L. Scharfetter, H.K. Gummel, "Large-signal analysis of a silicon Read 
diode oscillator,"
{\em{IEEE Trans. Electron Dev.}}, vol. 16, no.1, pp. 64-77, Jan. 1969.
\bibitem{10}
T. Tang, "Extension of the Scharfetter-Gummel algorithm to the energy
balance equation," {\em{IEEE Trans. Electron Dev. 1984}}, vol. 31,
no. 12, pp. 1912-1914, Dec. 1984.
\end{thebibliography}
\end{document}